\documentclass[dvips,11pt]{article} 
\usepackage{varioref,exscale,latexsym,amsmath,amssymb}
\usepackage{graphicx}
\usepackage{feynmp} 
\usepackage{mathrsfs}
\def\beq{\begin{equation}}

\def\eeq{\end{equation}}

\def\beqa{\begin{eqnarray}}

\def\eeqa{\end{eqnarray}}

\def\MeV{\mbox{ MeV}}

\def\GeV{\mbox{ GeV}}

\DeclareMathOperator{\Tr}{Tr}

\title{{\bf Long distance effects and strangeness in the nucleon}}
\author{John F. Donoghue$^a$, Barry R. Holstein$^a$,\\
Tobias Huber$^{a,b,c}$ and Andreas Ross$^{a,b}$\\ \\
$^a$ Department of Physics-LGRT\\
University of Massachusetts\\
Amherst, MA  01003\\  \\
$^b$ Institut f\"ur Theoretische Teilchenphysik\\
Universit\"at Karlsruhe\\
Karlsruhe, Germany\\  \\
$^c$ Institut f\"ur Theoretische Physik\\
Universit\"at Z\"urich\\
Z\"urich, Switzerland}
\begin{document}
\begin{titlepage}
\maketitle
\begin{abstract}
We discuss the calculation of the strange magnetic radius of the
proton in chiral perturbation theory. In particular we investigate
the low energy component of the loop integrals involving kaons. We
separate the chiral calculation into a low energy part and a high
energy component through use of a momentum space separation scale. This
separation shows that most of the chiral calculation comes from
high energies where the effective field theory treatment is not
valid. The resulting low energy prediction is in better agreement
with dispersive treatments. Finally, we
briefly discuss magnetic moments and show how our techniques can
help resolve an old puzzle in understanding the magnetic moments
of the proton and \(\Sigma^+\).
\end{abstract}
\setcounter{page}{0}
\end{titlepage}
\section{Introduction}

In this paper, we re-examine the issue of strangeness
components in the nucleon wavefunction.  Although the
valence quarks in the neutron and proton are non-strange, the
nucleons can acquire matrix elements of strange currents through
virtual effects. One traditional way to estimate such effects is
through kaon loops in the framework of chiral perturbation theory.
At first sight this seems quite appealing as chiral perturbation
theory provides a rigorous treatment of the Goldstone bosons as an
expansion in the energy and the masses of the quarks. However, it
is becoming clearer that the strange quark mass (and hence the
kaon mass) is too large for a reliable effective field theory
treatment in the baryon sector. In this paper we address one of
the cleanest measures of strangeness---the strange magnetic
radius---as well as some related quantities, and we discuss the
reliability of the effective field theory treatment.

The strange magnetic radius is one of those special quantities in
chiral perturbation theory that is finite at one loop without
requiring any contributions from low energy constants in the
chiral Lagrangian. The one loop result then represents a unique
parameter-free contribution\cite{hemm}. However, the chiral
prediction is in strong contradiction with the corresponding
dispersive calculation, which finds a much smaller
answer\cite{muso}. Part of the motivation for the present work
is to attempt to elucidate this discrepancy.  On the experimental
side the only experiment which focuses on the strange magnetic
form factor is SAMPLE, which was performed at MIT-Bates at a
momentum transfer $q^2=-0.1$ GeV$^2$~\cite{samp}.  There exist also
forward experiments---HAPPEX~\cite{happ} and G0~\cite{g0} at JLab
and PVA4 at Mainz~\cite{pva4}---which measure a linear combination
of strange magnetic and charge radius effects, but the statistical
precision is not yet sufficient to produce a meaningful
experimental value for the slope of strange magnetism, so our
conclusions will be based only on theoretical calculation.

 An effective field theory is
a technique for exploring the long-distance/low-energy predictions
of a more complete full theory. For mesonic chiral perturbation
theory, the separation between low-energy and high-energy often
occurs around 700 MeV, when the rho meson becomes important. In
the dispersive approach, the \(K \bar K\) cut in the t-channel
starts at an energy \(s = 4 m^2_K \sim 1 \GeV^2\). For scales
below this separation energy the effective field theory
description can be taken as reliable, while above this value new
degrees of freedom come into play and the effective treatment can
no longer be trusted.  This leads us to question how much of the
chiral kaon loop actually comes from the long distance regime and
is therefore reliably predicted.  We will address this issue
specifically below and show that almost all of the chiral
prediction for the strange magnetic radius comes from short
distances, where the effective field theory is no longer
believable.

\section{Notation}
The calculation of baryon magnetic moments and the strange magnetic radius requires the chiral Lagrangians given below, where we
follow closely~\cite{hemm}.
\beqa
\mathscr L_{M \! M} & = & \frac {F_\phi^{\, \, 2}} {4} \, \Tr \! \left( D_\mu U (D^\mu U)^\dagger \right)
    + \frac {F_\phi^{\, \, 2}} {4} \, \Tr \! \left(  \chi U^\dagger + U \chi^\dagger \right)\\
\displaystyle \mathscr L_{MB} & = & \Tr \! \left( \bar B \, i v \cdot D B \right)+ D \, \Tr \! \left( \bar B S^\mu \left\{u_\mu ,
B\right\}\right) + F \, \Tr \! \left( \bar B S^\mu \left[u_\mu , B\right]\right)
\eeqa
with
\beqa
 D_\mu U  & = & \partial_\mu U - i \left[v_\mu^{\left(i\right)}, U \right] + \ldots \\
 U  = \, u^2 & = & \mbox{ exp} \left(i \, \sqrt{2} \, \Phi/F_{\phi}\right) \\
 D_\mu B  & = & \partial_\mu B + \left[\Gamma_\mu, B\right] - i \, B  \Tr \left(v_\mu^{(0)}\right)\\
 \Gamma_\mu & = & \frac {1} {2} \, [u^\dagger , \partial_\mu u]
    - \frac {i} {2} u^\dagger \, v_\mu^{(i)} \, u - \frac {i} {2} u \,\, v_\mu^{(i)} \, u^\dagger + \ldots \\
 u_\mu & = & i u^\dagger \left(D_\mu U\right)  u^\dagger
\eeqa
\noindent Furthermore, we use $D = 3/4$ and $F = 1/2$ for the meson-baryon couplings and $F_\phi = (F_\pi+F_K)/2 \simeq 102 \MeV$ for the average
pseudoscalar decay constant. $\Tr$ denotes the trace in flavor space and $S_\mu\!\!~=~\!\!\textstyle \frac{i}{2} \, \gamma_5 \,
\sigma_{\mu\nu} \, v^{\nu}$ is the Pauli-Lubanski spin vector~\cite{spinvec}. $v_\mu^{(0)}$ stands for the singlet current whereas $v_{\mu}^{(i)}, \, i=1
\ldots 8,\) stands for the \(i-\)th component of the octet current. The meson and baryon fields are contained in the two matrices
\beq
\begin{array}{cc}
\Phi = \left(
     \begin{array}{*{3}{c}}
        \frac {\eta} {\sqrt{6}} + \frac{\pi^0}{\sqrt{2}} &   \pi^+  &  \textstyle K^+  \\
        \pi^-  &   \frac {\eta} {\sqrt{6}} - \frac{\pi^0}{\sqrt{2}} &  K^0  \\
         K^-  &  \overline{K}^0  &  - \frac {2 \, \eta} {\sqrt{6}}  \\
     \end{array}
     \right)& \hspace{-5pt}B  =  \left(
     \begin{array}{*{3}{c}}
         \frac {\Lambda} {\sqrt{6}} + \frac{\Sigma^0}{\sqrt{2}} &  \Sigma^+  &  p  \\
        \Sigma^-  &  \frac {\Lambda} {\sqrt{6}} - \frac{\Sigma^0}{\sqrt{2}}    &  n  \\
         \Xi^-  &  \Xi^0  &  - \frac {2 \, \Lambda} {\sqrt{6}}  \\
     \end{array}
     \right)
\end{array}
\eeq

In order to obtain baryon magnetic moments and the strange magnetic radius one studies vector current matrix elements in SU(3) HBChPT
according to
 \begin{equation}\label{def_sachs_ff}
 J^{(i)}_{\mu} = \frac{1}{N_1 N_2} \; \bar u \left(p^\prime\right) P^+_v \left[v_{\mu} \, G_E^{(i)} (q^2) +
    \left[S_\mu , S_\nu \right] \frac {q^\nu} {m} \, G_M^{(i)} (q^2)\right] P^+_v u\left(p\right)
 \end{equation}
 where \(\displaystyle q^2 = \left(p' - p\right)^2\) is the invariant momentum transfer squared and $m~=~1151 \MeV$ is the average mass
 of the baryon octet. The quantities \(\displaystyle G_{E,M}^{(i)} (q^2)\) are the  electric and magnetic Sachs form factors. Their
 relation to the Dirac and Pauli form factors $F_1^{(i)}(q^2)$ and $F_2^{(i)}(q^2)$ can be found in~\cite{ff}. The
 superscript \(i\) denotes the type of current one is interested in:
 \beqa
  \displaystyle J^{\, \mbox{\scriptsize (em)}}_{\mu} & := & \! \big< B \, | \, \textstyle{\frac {2} {3}} \;
  \displaystyle \bar u \, \gamma_{\mu} u - \textstyle{\frac {1} {3}} \; \displaystyle \bar d \, \gamma_{\mu}
  d - \textstyle{\frac {1} {3}} \; \displaystyle \bar s \, \gamma_{\mu}  s \, | \, B \big>
  \displaystyle = \textstyle{\frac {1} {2}} \; \displaystyle J^{(3)}_{\mu} + \textstyle{\frac {1} {2
  \sqrt{3}}} \; \displaystyle J^{(8)}_{\mu} , \, \\
 \displaystyle J^{\, (s)}_{\mu} & := & \! \big< B \, | \, \displaystyle \bar s \, \gamma_{\mu}  s \, | \, B \big>
  \displaystyle = \textstyle{\frac {1} {3}} \; \displaystyle J^{(0)}_{\mu} - \textstyle{\frac {1}
  {\sqrt{3}}} \; \displaystyle  J^{(8)}_{\mu}.
 \eeqa

From an experimental point of view, in addition to the electromagnetic
matrix elements one considers the neutral current which couples to the $Z^0$ and which is of the form
\beq
\displaystyle J_\mu^{(Z)} :=  \big< B \, | \, \textstyle \bar{u}_L \gamma_\mu u_L - \bar{d}_L\gamma_\mu d_L - 2\sin^2\theta_w \,
(\frac{2}{3} \; \bar{u}\gamma_\mu u-\frac{1}{3} \; \bar{d}\gamma_\mu d-\frac{1}{3} \; \bar{s}\gamma_\mu s) \, | \, B \big> \; .
\eeq
Since there are three active degrees of freedom, by combining
electromagnetic measurements on the neutron and proton with
parity-violating electron-proton scattering results one can
isolate the matrix element of the strange
quark --- $<B|\bar{s}\gamma_\mu s|B>$ --- and determine the
corresponding strangeness form factors.

The magnetic Sachs form factors have the following Taylor expansion around $q^2=0$ from which one can immediately extract magnetic moments
and the strange magnetic radius:
\beqa
\displaystyle  G^{\mbox{\scriptsize (em)}}_{M}(q^2) & = & \mu^{\mbox{\scriptsize (em)}} + \mathscr  O(q^2)\\
\displaystyle  G^{(s)}_{M}(q^2) & = & \mu^{(s)} + \frac {1} {6} \, \big<r^2_{M,s} \big> \cdot q^2 + \mathscr O(q^4).
\eeqa

As an extension of our examination of baryon properties we include intermediate states of the baryon decuplet. The dynamics of the
decuplet is contained in the Lagrangian~\cite{deculag1,deculag2}
\beq\label{deculag}
\mathscr L  = - \bar T^\mu \, i v \cdot D \, T_\mu + \frac {C} {2} \left(\bar T^\mu u_\mu B + \bar B u_\mu T^\mu\right) + H \, \, \bar
T^\mu S_\nu u^\nu T_\mu + \Delta \bar T ^\mu T_\mu
\eeq
with
\beqa
 D_\nu T^\mu_{abc} \!\!\!\!& = & \!\!\!\! \partial_\nu T^\mu_{abc} + \left(\Gamma_\nu\right)^d_a T^\mu_{dbc} +
 \left(\Gamma_\nu\right)^d_b T^\mu_{adc} + \left(\Gamma_\nu\right)^d_c
    T^\mu_{abd} - \, i \, T^\mu_{abc} \, \Tr (v_{\nu}^{(0)}) \quad \quad\\
   \bar T A B \!\!\!\!& = & \!\!\!\! \bar T_{jkl} \, A^{j}_{\; \, \,  m} \, B^{k}_{\; \, \, n} \, \epsilon^{lmn} \, ,  \quad
    \bar B A T  =  \bar B^{n}_{\; \, \, k} \, A^{m}_{\; \, \, j} \, T^{jkl} \, \epsilon_{\, lmn} \, , \\
    \bar T A T \!\!\!\!& = & \!\!\!\! \bar T_{jkl} \, A^{j}_{\; \, \, m} \; T^{mkl}.
\eeqa

\noindent The fields $T^\mu_{abc}$ are totally symmetric in the indices $a$, $b$, and $c$. Their relation to the physical
states can be found in ref.~\cite{Ttostates}.
The quantity $\Delta$ that arises in the last term of eqn.~(\ref{deculag}) is the difference between the average
mass of the decuplet($m=1382$~MeV) and the average mass of the octet($m=1151$~MeV), hence $\Delta = 231 \MeV$. The value for the octet-decuplet coupling is taken to
be $C = -3/2$, see ref.~\cite{Borasoy}. Having set our formalism and notation,
we now proceed to our calculation.

\section{Chiral Corrections to $<r_{M,s}^2>$}
We have repeated the calculation of the strange magnetic form
factor of the proton~\cite{hemm}. In order to address the issue of
long vs. short distance contributions we performed the
calculations utilizing a momentum space regulator in the loop
integral. This technique, often called long distance
regularization, has been explored previously in \cite{ldr}. When
the renormalization of the low energy parameters is properly
performed, this method reproduces exactly the results of
dimensional regularization when the meson masses are small, or
equivalently in the limit that the cutoff parameter $\Lambda$ is
taken to infinity. However at finite $\Lambda$, the method only
admits long distance contributions ---{\it i.e.}, those with
$\delta r > 1/\Lambda$ --- since shorter distance contributions are
excised by the cutoff. This procedure then provides a diagnostic
of how much of the final result comes from long distance physics.

\begin{figure}[h]
 \begin{fmffile}{fmf2}
  \begin{equation*}
   \begin{fmfgraph*}(120,100)
    \fmfleft{Nin}
    \fmfright{Nout}
    \fmftop{gamma}
    \fmf{plain}{Nin,vul}
    \fmf{plain}{vul,vur}
    \fmf{plain}{vur,Nout}
    \fmffreeze
    \fmf{phantom,label=(a),label.dist=20pt}{vul,vur}
    \fmf{dashes,left}{vul,vur}
    \fmfposition
    \fmfipath{p}
    \fmfiset{p}{vpath(__vul,__vur)}
    \fmfiv{d.sh=circle,d.siz=2thick}{point length(p)/2 of p}
    \fmffreeze
    \fmfi{photon}{vloc(__gamma) .. point length(p)/2 of p}
    \fmfdot{vul,vur}
   \end{fmfgraph*}
   \hspace{25pt}
   \begin{fmfgraph*}(120,100)
    \fmfleft{Nin}
    \fmfright{Nout}
    \fmftop{gamma}
    \fmf{plain}{Nin,vul}
    \fmf{dbl_plain}{vul,vur}
    \fmf{plain}{vur,Nout}
    \fmffreeze
    \fmf{phantom,label=(b),label.dist=20pt}{vul,vur}
    \fmf{dashes,left}{vul,vur}
    \fmfposition
    \fmfipath{p}
    \fmfiset{p}{vpath(__vul,__vur)}
    \fmfiv{d.sh=circle,d.siz=2thick}{point length(p)/2 of p}
    \fmffreeze
    \fmfi{photon}{vloc(__gamma) .. point length(p)/2 of p}
    \fmfdot{vul,vur}
   \end{fmfgraph*}
   {}\vspace*{-30pt}
  \end{equation*}
 \end{fmffile}
 \caption{Feynman diagrams contributing to the strange magnetic radius at $\mathcal O(p^3)$.
Solid lines are octet baryons, dashed lines are pseudogoldstone bosons and double lines are
decuplet baryons.\label{feynmdiagrams}}
\end{figure}
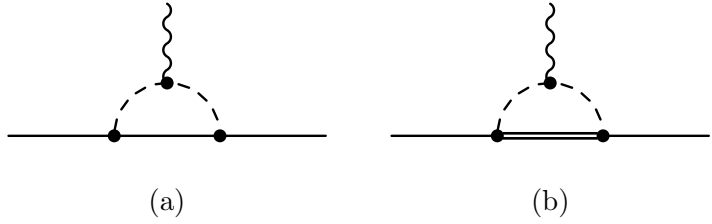

The strange magnetic radius is defined by $\displaystyle
\big<r^2_{M,s} \big> = 6 \cdot \mbox{d}G^{(s)}_M(q^2) /
 \mbox{d} q^2 \big|_{q^2=0}$ and
 there exists a single diagram with octet baryons and pseudo-Goldstone
bosons as intermediate states that contributes to it---{\it cf}.
Figure \ref{feynmdiagrams}a. When one also includes decuplet
baryons as intermediate states there exists an additional diagram
as shown in Figure \ref{feynmdiagrams}b. Throughout, we will
work in the Breit frame where $v\cdot q =0$\cite{ff}.

For both diagrams there is a single generic loop integral which
contributes to the strange magnetic form factor at order $\mathcal
O(p^3)$. Before regularization it reads
\begin{equation}\label{integraldef}
  i \! \! \int \! \frac {d^4 k} {\left(2 \pi\right)^4}
   \frac {k_\mu k_\nu} {(- v \! \cdot \! k \! - \! \Delta \! +\!  i \epsilon) (k^2 \! - \! M^2 \! + \! i \epsilon) (\!(k\! +\! q)^2 \! \! - \! M^2 \! + \! i \epsilon)}
  \! = \! I(M,q^2\!,\Delta) g_{\mu\nu} + \dots
\end{equation}
and the only term contributing to the strange magnetic form factor
is the piece proportional to \(g_{\mu \nu}\). The parameter
$\Delta=M_\Delta-M_N$ is non-zero only for the diagram with
decuplet intermediate states. Since we are primarily discussing
the strange magnetic radius we expand \( I(M,q^2\!,\Delta)\) in
powers of \(q^2\)---
\begin{equation}
  I(M,q^2\!,\Delta) =  I_0(M,\Delta) +  I_1(M,\Delta) \, q^2 + \dots
\end{equation}
Then \( I_1(M,\Delta)\), the part proportional to \(q^2\), is the piece
that enters the calculation of \(\displaystyle \big<r^2_{M,s}
\big>\).

For simplicity let us first consider only octet baryons as
intermediate states so that \(\Delta = 0\). In dimensional
regularization we find
\begin{equation}
 I_1^{d.r.}(M,0) = \frac {1} {16 \pi^2} \left(-\frac {\pi} {12 M}\right) = - \frac {1} {192 \pi M}
\end{equation}
When applying a dipole regulator for the diagrams in Fig.
\ref{feynmdiagrams} we associate a monopole form factor with each
internal meson line with the respective momenta \(k\) and
\(k+q\)---{\it i.e.}, we regularize the integral of Eq.
(\ref{integraldef}) by multiplying it by a factor
\begin{equation}
 \left(\frac {-\Lambda^2} {k^2-\Lambda^2+i\epsilon} \right) \left(\frac {-\Lambda^2} {(k+q)^2-\Lambda^2+i\epsilon}
 \right),
\end{equation}
obtaining
\begin{equation}
  I_1^{\Lambda}(M,0) =  I_1^{d.r.}(M,0) \cdot X \! \textstyle{\big(\frac {M} {\Lambda}\big)} = \displaystyle{- \frac {1} {192 \pi M}} \cdot X \! \textstyle{\big(\frac {M} {\Lambda}\big)}
\end{equation}
with
\begin{equation}
 X(x) = \frac {1 + \textstyle{\frac{14}{5}} \, x + x^2} {\left(1+x\right)^5}.
\end{equation}

Our result for the strange magnetic radius without decuplet
contributions is then
\begin{align}\label{strange_magnetic_radius_explicit}
  \displaystyle \big<r^2_{M,s} \big> & = - \frac {m \, (5 \, D^2 - 6 \, DF + 9 \, F^2)} {48 \, \pi \, M_K \,
  F_{\phi}^2} \cdot X \! \textstyle{\big(\frac {M_K} {\Lambda}\big)} \notag \\
  & = \big<r^2_{M,s,\mbox{\scriptsize dim.reg.}} \big> \cdot X \! \textstyle{\big(\frac {M_K}
  {\Lambda}\big)} = - \, 0.162 \mbox{ fm}^2 \cdot X \! \textstyle{\big(\frac {M_K}{\Lambda}\big)}.
 \end{align}
We observe then that integration over the loop in the presence
of the cutoff yields the factor \(X \! \textstyle{\big(\frac {M}
{\Lambda}\big)}\), which has the property \(X(0) = 1\). This means
that as the cutoff goes to infinity, or equivalently as the mass
gets small, we recover the dimensional regularization result.
Furthermore, the limit \(X(\infty) = 0\) expresses our expectation
that for infinitely heavy intermediate masses there is {\it no}
contribution, since such states decouple~\cite{app}.

\begin{figure}[h]
 \begin{center}
   \includegraphics[scale=0.8]{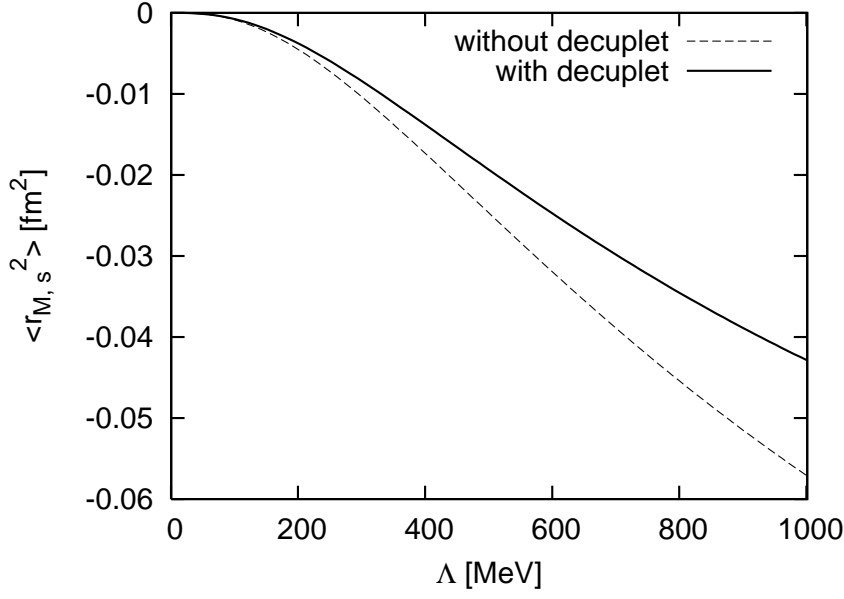}
  \end{center}
  \caption{Strange magnetic radius squared with (solid) and without (dashed) inclusion of decuplet loops.}
  \label{smrplot}
\end{figure}

This calculation allows us to estimate how much of the result
arises from long-distance physics. We do this by comparing the
answer for a given cutoff to the answer for infinite cutoff.  The
ratio of these two numbers yields the fraction of the dimensional
regularization result that comes from loop momenta below the
cutoff, or equivalently from distance scales larger than
$1/\Lambda$.  We exhibit the cutoff-dependence of the strange
magnetic radius in Fig. \ref{smrplot}, and quote the ratio to that
of dimensional regularization in Table 1. We observe that very
little of the dimensional regularization result comes from long
distance scales.  Indeed, for a reasonable cutoff value of 600
MeV, less than 20\% of the dimensional regularization result is
obtained.  This result suggests that even though the strange
magnetic radius is uniquely predicted in ${\cal O}(p^3)$ heavy
baryon chiral perturbation theory, using the physical value of the
kaon mass, most of the dimensionally regularized result comes from
distance scales so small that the effective field theory is not
believable.
\begin{table}[h]
 \begin{center}
  \begin{tabular}{|c|c|c|c|c|c|c|} \hline
   \(\Lambda\)/MeV & \(300\)    & \(400\)    & \(500\)    & \(600\)    & \(700\) & \(1000\)\\ \hline
   \(\displaystyle \big< r^2_{M,s} \big>_8\) / fm$^2$ & \(-0.010\) & \(-0.017\) &
   \(-0.025\) & \(-0.032\) & \(-0.039\) & \(-0.057\)\\ \hline
   \(\displaystyle \frac{\big< r^2_{M,s} \big>_8}{\big< r^2_{M,s, \mbox{\tiny d.r.}} \big>_8} \)
    & \(0.064\)  & \(0.107\)  & \(0.152\)  & \(0.197\)  & \(0.240\) & \(0.352\) \\ \hline
   \(\displaystyle \big< r^2_{M,s} \big>_{8+10}\) / fm$^2$ & \(-0.008\) & \(-0.014\) &
   \(-0.019\) & \(-0.025\) & \(-0.030\) & \(-0.043\)\\ \hline
   \(\displaystyle \frac{\big< r^2_{M,s} \big>_{8+10}}{\big< r^2_{M,s, \mbox{\tiny d.r.}} \big>_8} \)
    & \(0.052\)  & \(0.085\)  & \(0.119\)  & \(0.153\)  & \(0.184\) & \(0.264\) \\
     \hline
  \end{tabular}
  \caption{$\big< r^2_{M,s} \big>$ for different values of $ \Lambda $ using octet only and octet plus decuplet
  intermediate states.}
  \label{ratios}
 \end{center}
\end{table}

\newpage
We have also studied the effects of including the decuplet
intermediate states and find
\begin{equation}\label{rmsplusdec}
  \displaystyle \big<r^2_{M,s} \big> = \big<r^2_{M,s} \big>^{octet} + \frac {2 m C^2}
  {(4 \pi F_\phi)^2} \,  A^{(1)} \left(M_K, \Lambda, \Delta \right)
\end{equation}
with the octet part \(\displaystyle \big<r^2_{M,s} \big>^{octet}\)
as given in Eq. (\ref{strange_magnetic_radius_explicit}) and the
decuplet loop function
\begin{align}
 A&^{(1)} (M, \Lambda, \Delta) = \notag \\
\Lambda^4 & \Bigg[
\frac {\Delta (-8 \Delta^2 \! + 7 M^2 \! + 7 \Lambda^2)} {15 \left(M^2-\Lambda^2\right)^4} - \frac {2 \Delta (8 \Delta^4 \! \hspace{1pt} + \! \hspace{1pt} 15 M^2 \Lambda^2 \! \hspace{1pt} - \! \hspace{1pt} 10 \Delta^2 \! \hspace{1pt} (M^2 \! + \! \hspace{1pt} \Lambda^2))} {15 \left(M^2-\Lambda^2\right)^5} \, \ln \frac {M^2} {\Lambda^2} \notag \\
& {} \hspace*{-2.9pt} + \! \Bigg(\frac {1} {12 (M^2 \! \hspace{1pt} - \! \hspace{1pt} \Delta^2) \! \hspace{1pt} (M^2 \! \hspace{1pt} - \! \hspace{1pt} \Lambda^2)^2} - \frac {4 (M^2 \! \hspace{1pt} - \! \hspace{1pt} \Delta^2) \! \hspace{1pt} (4 \Delta^2 \! \hspace{1pt} + \! \hspace{1pt} M^2 \! \hspace{1pt} - \! \hspace{1pt} 5 \Lambda^2)} {15 \left(M^2-\Lambda^2\right)^5} \Bigg) F\left(M, \Delta\right) \notag \\
& {} \hspace*{-2.9pt} + \! \Bigg(\frac {1} {12 (\Lambda^2 \! \hspace{1pt} - \! \hspace{1pt} \Delta^2) \! \hspace{1pt} (M^2 \! \hspace{1pt} - \! \hspace{1pt} \Lambda^2)^2} + \frac {4 (\Lambda^2 \! \hspace{1pt} - \! \hspace{1pt} \Delta^2) \! \hspace{1pt} (4 \Delta^2 \! \hspace{1pt} + \! \hspace{1pt} \Lambda^2 \! \hspace{1pt} - \! \hspace{1pt} 5 M^2)} {15 \left(M^2-\Lambda^2\right)^5} \Bigg) F\left(\Lambda, \Delta\right) \Bigg]
\end{align}
where
\begin{equation}
     F\left(M , \Delta\right) =
    \begin{cases}
      \displaystyle
      \sqrt {M^2 - \Delta^2} \left(\pi - 2 \, \text{Arctan} \frac {\Delta} {\sqrt {M^2 - \Delta^2}}\right) & \text{for } |M| > |\Delta|\\\\
      \displaystyle
      \sqrt {\Delta^2 - M^2} \, \ln \frac {\Delta - \sqrt {\Delta^2 - M^2}} {\Delta + \sqrt {\Delta^2 - M^2}} & \text{for } |M| \le |\Delta|\\
    \end{cases}
\end{equation}
The resulting values for the magnetic strangeness radius are
quoted in Table~\ref{ratios}.

Let us discuss how our results are connected to other research in this topic.
We have highlighted the comparison to the
leading model-independent result, which counts as order $p^3$
in the chiral expansion\cite{hemm}. The long-distance portion of the one loop
integral includes not only the order $p^3$ result but also higher
order pieces, since the cut-off loop function contains terms that are
formally of all orders in the chiral expansion. When one works to higher order
in the chiral expansion, there will be low-energy constants that enter
at the next higher orders. These are regularization scheme dependent,
and so these constants will not be the same in dimensional regularization
and long-distance regularization. In dimensional regularization, the higher order
terms can correct for the misleadingly large leading result. Indeed this seems
to happen in a recent calculation performed to order $p^4$~\cite{Hammer}. Although there
is an unknown low-energy constant that enters at the next order, which leads to a large
uncertainty in the predicted value, the result that is obtained when this constant is set
equal to zero is only a quarter of the order $p^3 $ result. This is qualitatively similar
to our result, although our mechanism is not strictly of order $p^4$. It would be interesting
to carry out a full calculation to order $p^4$ using long-distance regularization throughout.
Our result is
even more directly connected to the dispersive treatment of the strange radius\cite{hemm, hammer2}.
The smallness of the dispersive result is related to the distant $K\bar{K}$ threshold, which
is similar to our suppression by the fact that the kaon mass is so heavy that very
little of the kaon loop integral is truly long-distance in character.

\section{Other aspects of the electromagnetic matrix elements}
We have also explored the full treatment of the loop corrections
to the electromagnetic currents of the octet of baryons. Details
can be found in~\cite{tt}. Here we report one of the interesting
results of this treatment---SU(3) breaking in the magnetic moments
of the baryons.

One of the early successes of the quark model was the
understanding of the magnetic moments of the baryons. Indeed, it
is relatively easy to understand the magnitude of the magnetic
moments of the neutron and proton in terms of the magnetic moments
of the quarks, as well as naturally predicting the ratio
$\mu_N/\mu_P = -2/3$. For the other baryons one introduces SU(3)
breaking by allowing the strange quark to have a different
magnetic moment due to its heavier mass (and hence smaller magnetic
moment). The \(\Lambda\) magnetic moment is a
particularly nice example of this, since in a valence quark model the
moment of the \(\Lambda\) is just the moment of the strange quark.
Because the strange quark is more massive than the up and down
quarks, the \(\Lambda\) magnetic moment is 30\% smaller than its
SU(3) prediction.

However, the quark model treatment of SU(3) breaking also has some
puzzling failures. This has been highlighted by
Lipkin~\cite{Lipkin} using the magnetic moment of the $\Sigma^+$.
He considers the ratio
\begin{equation}
R = \frac{\mu_{\Sigma^+} -\mu_P}{\frac13 (\mu_{\Xi^0} -
\mu_{\Xi^-})}
\end{equation}
for which the experimental value is
\begin{equation}
R_{exp}=1.68\pm 0.09
\end{equation}
In the quark model this ratio is rather directly related to the
quark moments
\begin{equation}
R_{QM} = \frac{\mu_{d} -\mu_s}{ \mu_{d}} \sim 0.3
\end{equation}
and it is difficult to modify this ratio in any significant way.
Lipkin explores this problem in detail - it seems to be a firm
prediction of the quark model. We will refer to this as the
$\Sigma^+$ puzzle. The resolution must come from physics that is
beyond the quark model.

It appears that chiral loops help to resolve the $\Sigma^+ $
puzzle. Chiral loop corrections to the magnetic moments have been
studied in the past. In an SU(3) chiral perturbation evaluation
using dimensional regularization, these loop effects are so large
that they essentially destroy the approximate understanding
provided by the quark model~\cite{Jenkins}. The corrections are so
big that the chiral expansion has broken down. Previous studies
have shown that the use of long-distance regularization, either
without~\cite{ldr} or with~\cite{Borasoy} decuplet fields, reduces
the magnitude of the corrections to a manageable level. However,
the effect on phenomenology of the long distance part of the loops
was not studied.  Here we note that the SU(3) breaking pattern of
chiral loops with long-distance regularization helps improve the
phenomenology of either SU(3) fits or quark model fits to the
moments.

First let us consider the description of the magnetic moments
within the chiral expansion. This is a purely symmetry-based
method. At tree level, the magnetic moments are described by two
SU(3) parameters, which can be called $b_D$ and $b_F$. As
described in~\cite{ldr}, the diverging powers of the cutoff
$\Lambda$ are absorbed into the renormalized values of these
parameters. The residual effects are contained in the loop
integrals. Long distance regularization has been developed for
this problem elsewhere~\cite{ldr,Borasoy} and we only present the
results. At this level of the chiral expansion, we are including
only the lowest order SU(3) parameters and the loop integrals. All
of the SU(3) breaking is then contained in the loops. Of course,
our fits could be further refined by including higher order
parameters in the chiral expansion. Such parameters are surely
present and would certainly improve the agreement with experiment
--- with some loss of predictive power. However, we are interested
in demonstrating the basic quality of the fit without invoking
these extra parameters.

We consider first the case where only the octet baryons are
treated in loop diagrams and then the case where the decuplet is
included. For the proton and the neutron, the  octet results are
\begin{eqnarray}\label{mu_em_p_ren}
\mu_p^{\mbox{\scriptsize ren}} =  &&1 + b_F^r + \frac {1} {3} \,
b_D^r + \frac {m} {24 \, \pi \, F_{\phi}^2} \times
\nonumber \\
   &&\left[\frac{\left(F+D\right)^2
\, \Lambda^4}{\left(\Lambda + M_{\pi}\right)^3} + \frac{2 \,
\left(D^2 + 3 \, F^2\right) \Lambda^4}{3 \, \left(\Lambda +
 M_K\right)^3} - \frac {\left(5 \, D^2 + 6 \, DF + 9 \, F^2\right)
\Lambda} {3}\right],\nonumber\\
\qquad
 \end{eqnarray}
and
 \begin{equation}\label{mu_em_n_ren}
 \mu_n^{\mbox{\scriptsize ren}} = - \frac {2} {3} \, b_D^r + \frac {m}
{24 \, \pi \, F_{\phi}^2} \left[-\frac{\left(F+D\right)^2 \,
\Lambda^4}{\left(\Lambda + M_{\pi}\right)^3} + \frac{\left(D-
F\right)^2 \, \Lambda^4}{\left(\Lambda + M_K\right)^3} + 4 \, DF
\,
  \Lambda\right],
 \end{equation}
The limit as $\Lambda \to \infty$ corresponds to the result of
dimensional regularization~\cite{emff}. The limit $\Lambda \to 0$ is
equivalent to not including the chiral loops at all. The
coefficients for the other amplitudes are well known and are given
in the references~\cite{Jenkins,ldr,Borasoy,emff}. In order to pin down
the low-energy constants \(b_D^r\) and \(b_F^r\) we perform a
 least-squares fit to the experimental data, such
 that the quantity
 \begin{equation}\label{chi_square}
\chi^2_{\sigma} := \frac {1} {N} \sum_{i} \left[\frac
{\mu_i^{\mbox{\scriptsize ren}} - \mu_i^{\mbox{\scriptsize exp}}}
{\mu_i^{\mbox{\scriptsize exp}} \cdot (\sigma/100)}\right]^2
 \end{equation}
 gets minimized. The sum over \(i\) covers all baryons in
Table~\ref{mag_mom_fit}, \(N\) is the total number of magnetic
 moments and \(\sigma\) is the allowed percental deviation from the
experimental value. The quantity \(\chi^2_{\sigma}\) then
 contains information about the quality of the fit. Small values of
\(\chi^2_{\sigma}\) (around unity and below) indicate that
 theory and experiment agree within the considered percental deviation.
In Table~\ref{mag_mom_fit} we show the result
of the fits performed for several values of
the separation scale $\Lambda$.
\begin{table}
\begin{center}
{}\hspace*{-1cm}
  \begin{tabular}{|c||c|c|c|c|c|c|c||c||c|}\hline
\small\( \Lambda / \mbox{MeV}\)                    & \(\mbox{\small 300}\)    & \(\mbox{\small 400}\)    & \(\mbox{\small 500}\)    & \(\mbox{\small 600}\)    & \(\mbox{\small 700}\)    & \(\mbox{\small 1000}\)   & \(\mbox{\small 1100}\)   &\small{dim. reg.}         & \small {exp.}            \\ \hline\hline
      \( b_D^r\)                                   & \(\mbox{\small 3.085}\)  & \(\mbox{\small 3.305}\)  & \(\mbox{\small 3.498}\)  & \(\mbox{\small 3.667}\)  & \(\mbox{\small 3.818}\)  & \(\mbox{\small 4.178}\)  & \(\mbox{\small 4.276}\)  & \(\mbox{\small 6.114}\)  &                          \\ \hline
      \( b_F^r\)                                   & \(\mbox{\small 1.187}\)  & \(\mbox{\small 1.349}\)  & \(\mbox{\small 1.492}\)  & \(\mbox{\small 1.617}\)  & \(\mbox{\small 1.728}\)  & \(\mbox{\small 1.994}\)  & \(\mbox{\small 2.065}\)  & \(\mbox{\small 3.416}\)  &                          \\ \hline\hline
      \(\mu_p^{\mbox{\scriptsize ren}}\)           & \(\mbox{\small 2.384}\)  & \(\mbox{\small 2.443}\)  & \(\mbox{\small 2.502}\)  & \(\mbox{\small 2.559}\)  & \(\mbox{\small 2.612}\)  & \(\mbox{\small 2.749}\)  & \(\mbox{\small 2.787}\)  & \(\mbox{\small 3.595}\)  & \(\mbox{\small 2.792847}\)  \\ \hline
      \(\mu_n^{\mbox{\scriptsize ren}}\)           & \(\mbox{\small -1.616}\) & \(\mbox{\small -1.696}\) & \(\mbox{\small -1.777}\) & \(\mbox{\small -1.854}\) & \(\mbox{\small -1.927}\) & \(\mbox{\small -2.113}\) & \(\mbox{\small -2.165}\) & \(\mbox{\small -3.263}\) & \(\mbox{\small -1.91304}\) \\ \hline
      \(\mu_{\Sigma^+}^{\mbox{\scriptsize ren}}\)  & \(\mbox{\small 2.303}\)  & \(\mbox{\small 2.313}\)  & \(\mbox{\small 2.323}\)  & \(\mbox{\small 2.333}\)  & \(\mbox{\small 2.342}\)  & \(\mbox{\small 2.366}\)  & \(\mbox{\small 2.373}\)  & \(\mbox{\small 2.513}\)  & \(\mbox{\small 2.458}\)  \\ \hline
\(\mu_{\Sigma^-}^{\mbox{\scriptsize ren}}\)        & \(\mbox{\small -0.871}\) & \(\mbox{\small -0.912}\) & \(\mbox{\small -0.953}\) & \(\mbox{\small -0.992}\) & \(\mbox{\small -1.029}\) &\(\mbox{\small -1.124}\)  & \(\mbox{\small -1.150}\) & \(\mbox{\small -1.710}\) & \(\mbox{\small -1.16}\)  \\ \hline
\(\mu_{\Lambda}^{\mbox{\scriptsize ren}}\)         & \(\mbox{\small -0.716}\) & \(\mbox{\small -0.701}\) & \(\mbox{\small -0.685}\) & \(\mbox{\small -0.671}\) & \(\mbox{\small -0.657}\) & \(\mbox{\small -0.621}\) & \(\mbox{\small -0.611}\) & \(\mbox{\small -0.401}\) & \(\mbox{\small -0.613}\) \\ \hline
\(\mu_{\Lambda\Sigma^0}^{\mbox{\scriptsize ren}}\) & \(\mbox{\small 1.342}\)  & \(\mbox{\small 1.377}\)  & \(\mbox{\small 1.412}\)  & \(\mbox{\small 1.446}\)  & \(\mbox{\small 1.477}\)  & \(\mbox{\small 1.558}\)  & \(\mbox{\small 1.581}\)  & \(\mbox{\small 2.057}\)  & \(\mbox{\small 1.61}\)   \\ \hline
\(\mu_{\Xi^0}^{\mbox{\scriptsize ren}}\)           & \(\mbox{\small -1.425}\) & \(\mbox{\small -1.389}\) & \(\mbox{\small -1.354}\) & \(\mbox{\small -1.320}\) & \(\mbox{\small -1.289}\) & \(\mbox{\small -1.207}\) & \(\mbox{\small -1.185}\) &\(\mbox{\small -0.704}\)  & \(\mbox{\small -1.25}\)  \\ \hline
\(\mu_{\Xi^-}^{\mbox{\scriptsize ren}}\)           & \(\mbox{\small -0.775}\) & \(\mbox{\small -0.758}\) & \(\mbox{\small -0.741}\) & \(\mbox{\small -0.725}\) & \(\mbox{\small -0.710}\) &\(\mbox{\small -0.671}\)  & \(\mbox{\small -0.660}\) & \(\mbox{\small -0.430}\) & \(\mbox{\small -0.6507}\) \\ \hline\hline
      \(\chi^2_{10}\)                              & \(\mbox{\small 2.795}\)  & \(\mbox{\small 1.987}\)  & \(\mbox{\small 1.327}\)  & \(\mbox{\small 0.837}\)  & \(\mbox{\small 0.504}\)  & \(\mbox{\small 0.211}\)  & \(\mbox{\small 0.274}\)  & \(\mbox{\small 16.332}\) &            \\ \hline
      \(\chi^2_{12}\)                              & \(\mbox{\small 1.941}\)  & \(\mbox{\small 1.380}\)  & \(\mbox{\small 0.922}\)  & \(\mbox{\small 0.581}\)  & \(\mbox{\small 0.350}\)  & \(\mbox{\small 0.146}\)  & \(\mbox{\small 0.190}\)  & \(\mbox{\small 11.342}\) &            \\ \hline
      \(\chi^2_{15}\)                              & \(\mbox{\small 1.242}\)  & \(\mbox{\small 0.883}\)  & \(\mbox{\small 0.590}\)  & \(\mbox{\small 0.372}\)  & \(\mbox{\small 0.224}\)  & \(\mbox{\small 0.094}\)  & \(\mbox{\small 0.122}\)  & \(\mbox{\small 7.259}\)  &            \\ \hline\hline
      \(R\)                                        & \(\mbox{\small 0.374}\)  & \(\mbox{\small 0.617}\)  & \(\mbox{\small 0.875}\)  & \(\mbox{\small 1.138}\)  & \(\mbox{\small 1.398}\)  & \(\mbox{\small 2.142}\)  & \(\mbox{\small 2.373}\)  & \(\mbox{\small 11.85}\)  & \(\mbox{\small 1.676}\)  \\ \hline
 \end{tabular} \end{center}
 \caption{Least-squares fit to the octet magnetic moments including octet intermediate states.
 As described in the text,
the factors of $\chi^2_\sigma$ describe the quality of the fit allowing a fractional uncertainty
of $\sigma\%$.
\label{mag_mom_fit}}
 \end{table}\\

We see that the fit is reasonable (better that 10\% for many
values of $\Lambda$) even without any other sources of SU(3)
breaking besides the chiral loops. However, the fit with
dimensional regularization is very poor. This quantifies the
message of~\cite{ldr}, that the dimensional regularization result contains a very
large short-distance component (corresponding to energies beyond
the scale $\Lambda$, which upsets the phenomenology of baryons).
This big short distance component requires large coefficients
from higher orders
in the chiral lagrangian if we are to restore a successful
phenomenology. In contrast, when only
the long-distance parts of loops are included, there is no need for any
additional significant
ingredients. We conclude that the chiral loops aid the phenomenology
for moderate values of $\Lambda$.

This conclusion is reinforced if we include decuplet baryons in
the chiral loops. Here again the dimensionally regularized result
leads to corrections that upset the phenomenology. Keeping only
the long distance portion of the loops, however, will yield modest
corrections that help the phenomenology. If we include the
decuplet states we have the additional modification of spin 3/2
loops, which include an additional loop function. This leads to the
structure:
\begin{eqnarray}\label{mu_em_octet_ren_decu}
\mu^{{ren}}_{{\rm tot}} &=& Q \, \left( 1+\tilde{b}_F^r\right) +
\alpha_D \,
\tilde{b}_D^r  \nonumber \\
&-& \frac {m} {24 \, \pi \, F_{\phi}^2} \left[
\beta_{\pi}\frac{\Lambda^4}{\left(\Lambda + M_{\pi}\right)^3} +
\beta_K \frac{\Lambda^4}{\left(\Lambda + M_K\right)^3} -
\left(\beta_{\pi} + \beta_K \right)
  \Lambda\right] \nonumber \\
  &+& \frac {m \, C^2} {(4 \pi F_{\phi})^2}
\left[\beta_{\pi}^d \cdot A^{(2)}(\Lambda,M_{\pi},\Delta) +
\beta_K^d
  \cdot A^{(2)}(\Lambda,M_{K},\Delta)\right.\nonumber\\
   &+&\left. (\beta_{\pi}^d + \beta_K^d)
\left[\textstyle{\frac {\pi} {3}} \displaystyle \cdot \Lambda - 2
\, \Delta \, \mbox{ln} \! \left(\textstyle{\frac {\Lambda}
   {M_K}}\right)\right]\right],
 \end{eqnarray}
We have renormalized the SU(3) parameters such that in the limit $\Lambda \to \infty$
the results correspond to dimensional
regularization at a scale $\mu = m_K$.
The appropriate coefficients are given below, in Table~\ref{betas}, and in refs.~\cite{Jenkins, emff}. One finds:
\beq
\displaystyle  \tilde{b}_F^r = b_F^r, \qquad \tilde{b}_D^r = b_D^r + \frac{m \, C^2} {(4 \pi F_{\phi})^2} \left[\frac {\pi} {3} \cdot
\Lambda - 2 \, \Delta \, \mbox{ln} \! \left(\textstyle{\frac {\Lambda}
   {M_K}}\right) \displaystyle \right],
\eeq
\beqa
A^{(2)}(\Lambda,M,\Delta) & := &\frac
 {\left(4 \, \Delta^2 \! - \! \Lambda^2 \! - \! 3 \, M^2\right) \mbox{F}(\Lambda,\Delta)} {3 \, \left(\Lambda^2 - M^2\right)^3} \cdot \Lambda^4  \nonumber\\
 &&  - \frac
 {\left(4 \, \Delta^2 \! - \! M^2 \! - \! 3
 \, \Lambda^2\right) \mbox{F}(M,\Delta)} {3 \, \left(\Lambda^2 - M^2\right)^3} \cdot \Lambda^4  \\
 \displaystyle && - \frac {2 \, \Delta \left\{ \Lambda^2 - M^2 + \left[4 \, \Delta^2 - 3 \left(\Lambda^2 + M^2\right)\right] \mbox{ln} \!
 \left(\frac  {\Lambda}{M}\right)\right\}} {3 \, \left(\Lambda^2 - M^2\right)^3} \cdot \Lambda^4. \nonumber
\eeqa
\begin{table}[t]
\begin{center}
{}\hspace*{-1cm}
 \begin{tabular}{c|c|c}
baryon                            & \(\displaystyle \beta_{\pi}^d\) & \(\displaystyle \beta_{K}^d\) \\ \hline
\(\displaystyle p\)               & \(- \, 4/9\)                    & \(1/9\)                       \\ \hline
\(\displaystyle n\)               & \(4/9\)                         & \(2/9\)                       \\ \hline
\(\displaystyle \Sigma^+\)        & \(1/9\)                         & \(-4/9\)                      \\ \hline
\(\displaystyle \Sigma^0\)        & \(0\)                           & \(- \, 1/3\)                  \\ \hline
\(\displaystyle\Sigma^-\)         & \(- \, 1/9\)                    & \(- \, 2/9\)                  \\ \hline
\(\displaystyle \Lambda\)         & \(0\)                           & \(1/3\)                       \\ \hline
\(\displaystyle \Lambda\Sigma^0\) & \(- \, 2/(3 \, \sqrt{3})\)      & \(- \, 1/(3 \, \sqrt{3})\)    \\ \hline
\(\displaystyle \Xi^0\)           & \(2/9\)                         & \(4/9\)                       \\ \hline
\(\displaystyle \Xi^-\)           & \(- \, 2/9\)                    & \(- \, 1/9\)                  \\
          \end{tabular} \end{center}
 \caption{Coefficients $\displaystyle \beta_{\pi}^d$ and $\displaystyle \beta_{K}^d$. \label{betas}}
          \end{table}

Again we attempt a least squares fit to the magnetic moments,
including the SU(3) invariant parametrization plus the chiral
loops. The results are shown in Table~\ref{decu_mag_mom_fit}. The
fits are slightly better than the case above. Again dimensional
regularization produces large and unwelcome corrections, which
would have to be corrected in higher order.  For reasonable values of the cutoff, the chiral loops
produce modest and welcome corrections, improving on a pure SU(3)
analysis, as shown in Table~\ref{decu_mag_mom_fit}.\\
\begin{table}
 \begin{center}
{}\hspace*{-1cm}
\begin{tabular}{|c||c|c|c|c|c|c|c||c||c|}\hline
\( \Lambda / \mbox{MeV}\)                          & \(\mbox{\small 300}\)    & \(\mbox{\small 400}\)    & \(\mbox{\small 500}\)    & \(\mbox{\small 600}\)    & \(\mbox{\small 700}\)    & \(\mbox{\small 1000}\)   & \(\mbox{\small 1100}\)   &\small{dim. reg.}         & \small {exp.}      \\ \hline\hline
\( \tilde{b}_D^r\)                                 & \(\mbox{\small 3.915}\)  & \(\mbox{\small 4.058}\)  & \(\mbox{\small 4.210}\)  & \(\mbox{\small 4.358}\)  & \(\mbox{\small 4.499}\)  & \(\mbox{\small 4.864}\)  & \(\mbox{\small 4.967}\)  & \(\mbox{\small 7.206}\)  &                     \\ \hline
\( \tilde{b}_F^r\)                                 & \(\mbox{\small 1.190}\)  & \(\mbox{\small 1.356}\)  & \(\mbox{\small 1.501}\)  & \(\mbox{\small 1.629}\)  & \(\mbox{\small 1.743}\)  & \(\mbox{\small 2.017}\)  & \(\mbox{\small 2.091}\)  & \(\mbox{\small 3.506}\)  &                     \\ \hline\hline
\(\mu_p^{\mbox{\scriptsize ren}}\)                 & \(\mbox{\small 2.403}\)  & \(\mbox{\small 2.478}\)  & \(\mbox{\small 2.553}\)  & \(\mbox{\small 2.627}\)  & \(\mbox{\small 2.696}\)  & \(\mbox{\small 2.878}\)  & \(\mbox{\small 2.930}\)  & \(\mbox{\small 4.089}\)  & \(\mbox{\small 2.792847}\) \\ \hline
\(\mu_n^{\mbox{\scriptsize ren}}\)                 & \(\mbox{\small -1.627}\) & \(\mbox{\small -1.715}\) & \(\mbox{\small -1.805}\) & \(\mbox{\small -1.891}\) & \(\mbox{\small -1.972}\) & \(\mbox{\small -2.183}\) & \(\mbox{\small -2.242}\) & \(\mbox{\small -3.530}\) & \(\mbox{\small -1.91304}\) \\ \hline
\(\mu_{\Sigma^+}^{\mbox{\scriptsize ren}}\)        & \(\mbox{\small 2.295}\)  & \(\mbox{\small 2.300}\)  & \(\mbox{\small 2.304}\)  & \(\mbox{\small 2.308}\)  & \(\mbox{\small 2.311}\)  & \(\mbox{\small 2.319}\)  & \(\mbox{\small 2.320}\)  & \(\mbox{\small 2.331}\)  & \(\mbox{\small 2.458}\)    \\ \hline
\(\mu_{\Sigma^-}^{\mbox{\scriptsize ren}}\)        & \(\mbox{\small -0.874}\) & \(\mbox{\small -0.918}\) & \(\mbox{\small -0.962}\) & \(\mbox{\small -1.005}\) & \(\mbox{\small -1.044}\) & \(\mbox{\small -1.148}\) & \(\mbox{\small -1.177}\) & \(\mbox{\small -1.802}\) & \(\mbox{\small -1.16}\)    \\ \hline
\(\mu_{\Lambda}^{\mbox{\scriptsize ren}}\)         & \(\mbox{\small -0.711}\) & \(\mbox{\small -0.691}\) & \(\mbox{\small -0.671}\) & \(\mbox{\small -0.652}\) & \(\mbox{\small -0.633}\) & \(\mbox{\small -0.585}\) & \(\mbox{\small -0.572}\) & \(\mbox{\small -0.265}\) & \(\mbox{\small -0.613}\)   \\ \hline
\(\mu_{\Lambda\Sigma^0}^{\mbox{\scriptsize ren}}\) & \(\mbox{\small 1.351}\)  & \(\mbox{\small 1.393}\)  & \(\mbox{\small 1.436}\)  & \(\mbox{\small 1.478}\)  & \(\mbox{\small 1.517}\)  & \(\mbox{\small 1.619}\)  & \(\mbox{\small 1.648}\)  & \(\mbox{\small 2.290}\)  & \(\mbox{\small 1.61}\)     \\ \hline
\(\mu_{\Xi^0}^{\mbox{\scriptsize ren}}\)           & \(\mbox{\small -1.425}\) & \(\mbox{\small -1.389}\) & \(\mbox{\small -1.354}\) & \(\mbox{\small -1.320}\) & \(\mbox{\small -1.288}\) & \(\mbox{\small -1.207}\) & \(\mbox{\small -1.184}\) & \(\mbox{\small -0.701}\) & \(\mbox{\small -1.25}\)    \\ \hline
\(\mu_{\Xi^-}^{\mbox{\scriptsize ren}}\)           & \(\mbox{\small -0.773}\) & \(\mbox{\small -0.755}\) & \(\mbox{\small -0.737}\) & \(\mbox{\small -0.719}\) & \(\mbox{\small -0.703}\) & \(\mbox{\small -0.660}\) & \(\mbox{\small -0.648}\) & \(\mbox{\small -0.387}\) & \(\mbox{\small -0.6507}\)  \\ \hline\hline
\(\chi^2_{10}\)                                    & \(\mbox{\small 2.660}\)  & \(\mbox{\small 1.794}\)  & \(\mbox{\small 1.108}\)  & \(\mbox{\small 0.629}\)  & \(\mbox{\small 0.342}\)  & \(\mbox{\small 0.344}\)  & \(\mbox{\small 0.541}\)  & \(\mbox{\small 26.194}\) &                     \\ \hline
\(\chi^2_{12}\)                                    & \(\mbox{\small 1.848}\)  & \(\mbox{\small 1.246}\)  & \(\mbox{\small 0.769}\)  & \(\mbox{\small 0.437}\)  & \(\mbox{\small 0.238}\)  & \(\mbox{\small 0.239}\)  & \(\mbox{\small 0.376}\)  & \(\mbox{\small 18.191}\) &                     \\ \hline
\(\chi^2_{15}\)                                    & \(\mbox{\small 1.182}\)  & \(\mbox{\small 0.797}\)  & \(\mbox{\small 0.492}\)  & \(\mbox{\small 0.280}\)  & \(\mbox{\small 0.152}\)  & \(\mbox{\small 0.153}\)  & \(\mbox{\small 0.241}\)  & \(\mbox{\small 11.642}\) &                     \\ \hline\hline
\(R\)                                              & \(\mbox{\small 0.498}\)  & \(\mbox{\small 0.839}\)  & \(\mbox{\small 1.209}\)  & \(\mbox{\small 1.590}\)  & \(\mbox{\small 1.971}\)  & \(\mbox{\small 3.069}\)  & \(\mbox{\small 3.413}\)  & \(\mbox{\small 16.785}\) &  \(\mbox{\small 1.676}\)    \\\hline
 \end{tabular}
\end{center}
 \caption{Least-square fit to the octet magnetic moments with inclusion
of decuplet loops.\label{decu_mag_mom_fit}}
 \end{table}

Now let us turn to the quark model. In a pure valence quark model, the
only ingredients are the quark magnetic moments - the hadron
magnetic moments are sums of those of the quarks. The appropriate
linear combinations are given in many places, such as Table XII-2
of~\cite{dgh}. When we add chiral loops the issue is less
straightforward. Here the idea is that quarks provide a model for the short
distance physics, and we supplement the quark moments with the effects of
the long distance portions of chiral loops. In doing this, we will keep
the entire content of the loops - in particular we will not absorb the terms
which are linear in $\Lambda$ into definitions of renormalized parameters
as we did when we were implementing the SU(3) based fits. In practical terms then,
the quark moments replace the SU(3) parameters in the expressions given
above, and we drop the last terms in the expression of Eq 32,33 and 35 as these
were introduced in the renormalization procedure. The important new ingredient
is that there is now two sources of SU(3) breaking, that of the chiral loops and
the different moments of the quarks.

In this case we explore the phenomenology by fitting the proton,
neutron and $\Lambda$ moments exactly in order to determine the
$u,d,s$ magnetic moments, and then looking at the predictions for
the other baryon moments. The results are given in Table~\ref{quark_octet} and
Table~\ref{quark_decuplet}  for the cases of octet intermediate states only and for
octet and decuplet intermediate states.
 \begin{table}
 \begin{center}
  \begin{tabular}{|c|c|c|c|c|c|c|c|c||c||c}\hline
\small\( \Lambda  [\mbox{\footnotesize MeV}]\)     & $\mbox{\small tree}$   &\(\mbox{\small 300}\)    & \(\mbox{\small 400}\)    & \(\mbox{\small 500}\)    & \(\mbox{\small 600}\)    \\ \hline\hline
      \( \mu_u\)                                   & $\mbox{\small 1.852}$  &\(\mbox{\small 1.702}\)  & \(\mbox{\small 1.588}\)  & \(\mbox{\small 1.454}\)  & \(\mbox{\small 1.305}\)  \\ \hline
      \( \mu_d\)                                   & $\mbox{\small -0.972}$ &\(\mbox{\small -0.845}\) & \(\mbox{\small -0.758}\) & \(\mbox{\small -0.662}\) & \(\mbox{\small -0.561}\) \\ \hline
      \( \mu_s\)                                   & $\mbox{\small -0.613}$ &\(\mbox{\small -0.595}\) & \(\mbox{\small -0.574}\) & \(\mbox{\small -0.543}\) & \(\mbox{\small -0.504}\) \\ \hline\hline
      \(\mu_p\)                                    & $\mbox{\small 2.793}$  &\(\mbox{\small 2.793}\)  & \(\mbox{\small 2.793}\)  & \(\mbox{\small 2.793}\)  & \(\mbox{\small 2.793}\)  \\ \hline
      \(\mu_n\)                                    & $\mbox{\small -1.913}$ &\(\mbox{\small -1.913}\) & \(\mbox{\small -1.913}\) & \(\mbox{\small -1.913}\) & \(\mbox{\small -1.913}\) \\ \hline
      \(\mu_{\Lambda}\)                            & $\mbox{\small -0.613}$ &\(\mbox{\small -0.613}\) & \(\mbox{\small -0.613}\) & \(\mbox{\small -0.613}\) & \(\mbox{\small -0.613}\) \\ \hline
\(\mu_{\Sigma^+}\)                                 & $\mbox{\small 2.673}$  &\(\mbox{\small 2.629}\)  & \(\mbox{\small 2.602}\)  & \(\mbox{\small 2.575}\)  & \(\mbox{\small 2.548}\)  \\ \hline
\(\mu_{\Lambda\Sigma^0}\)                          & $\mbox{\small 1.630}$  &\(\mbox{\small 1.603}\)  & \(\mbox{\small 1.586}\)  & \(\mbox{\small 1.568}\)  & \(\mbox{\small 1.550}\)  \\ \hline
\(\mu_{\Sigma^-}\)                                 & $\mbox{\small -1.091}$ &\(\mbox{\small -1.053}\) & \(\mbox{\small -1.033}\) & \(\mbox{\small -1.016}\) & \(\mbox{\small -1.003}\) \\ \hline
\(\mu_{\Xi^0}\)                                    & $\mbox{\small -1.435}$ &\(\mbox{\small -1.389}\) & \(\mbox{\small -1.361}\) & \(\mbox{\small -1.332}\) & \(\mbox{\small -1.304}\) \\ \hline
\(\mu_{\Xi^-}\)                                    & $\mbox{\small -0.493}$ &\(\mbox{\small -0.542}\) & \(\mbox{\small -0.573}\) & \(\mbox{\small -0.607}\) & \(\mbox{\small -0.642}\) \\ \hline\hline
\(R\)                                              & $\mbox{\small 0.381}$  &\(\mbox{\small 0.581}\)  & \(\mbox{\small 0.728}\)  & \(\mbox{\small 0.903}\)  & \(\mbox{\small 1.107}\)  \\ \hline
 \end{tabular}
\end{center}

\begin{center}
   \begin{tabular}{|c|c|c|c|c|c|c|c|c||c||c}\hline
\small\( \Lambda  [\mbox{\footnotesize MeV}]\)     & $\mbox{\small 700}$    &\(\mbox{\small 800}\)    & \(\mbox{\small 900}\)    & \(\mbox{\small 1000}\)   &   \small {exp.}          \\ \hline\hline
      \( \mu_u\)                                   & $\mbox{\small 1.143}$  &\(\mbox{\small 0.972}\)  & \(\mbox{\small 0.792}\)  & \(\mbox{\small 0.605}\)  & \(\mbox{\small }\)       \\ \hline
      \( \mu_d\)                                   & $\mbox{\small -0.457}$ &\(\mbox{\small -0.351}\) & \(\mbox{\small -0.245}\) & \(\mbox{\small -0.137}\) & \(\mbox{\small }\)       \\ \hline
      \( \mu_s\)                                   & $\mbox{\small -0.458}$ &\(\mbox{\small -0.405}\) & \(\mbox{\small -0.347}\) & \(\mbox{\small -0.284}\) & \(\mbox{\small }\)       \\ \hline\hline
      \(\mu_p\)                                    & $\mbox{\small 2.793}$  &\(\mbox{\small 2.793}\)  & \(\mbox{\small 2.793}\)  & \(\mbox{\small 2.793}\)  & \(\mbox{\small 2.793}\)  \\ \hline
      \(\mu_n\)                                    & $\mbox{\small -1.913}$ &\(\mbox{\small -1.913}\) & \(\mbox{\small -1.913}\) & \(\mbox{\small -1.913}\) & \(\mbox{\small -1.913}\) \\ \hline
      \(\mu_{\Lambda}\)                            & $\mbox{\small -0.613}$ &\(\mbox{\small -0.613}\) & \(\mbox{\small -0.613}\) & \(\mbox{\small -0.613}\) & \(\mbox{\small -0.613}\) \\ \hline
\(\mu_{\Sigma^+}\)                                 & $\mbox{\small 2.524}$  &\(\mbox{\small 2.500}\)  & \(\mbox{\small 2.479}\)  & \(\mbox{\small 2.459}\)  & \(\mbox{\small 2.458}\)  \\ \hline
\(\mu_{\Lambda\Sigma^0}\)                          & $\mbox{\small 1.532}$  &\(\mbox{\small 1.514}\)  & \(\mbox{\small 1.498}\)  & \(\mbox{\small 1.481}\)  & \(\mbox{\small 1.61}\)   \\ \hline
\(\mu_{\Sigma^-}\)                                 & $\mbox{\small -0.993}$ &\(\mbox{\small -0.988}\) & \(\mbox{\small -0.986}\) & \(\mbox{\small -0.987}\) & \(\mbox{\small -1.16}\)  \\ \hline
\(\mu_{\Xi^0}\)                                    & $\mbox{\small -1.277}$ &\(\mbox{\small -1.251}\) & \(\mbox{\small -1.227}\) & \(\mbox{\small -1.204}\) & \(\mbox{\small -1.25}\)  \\ \hline
\(\mu_{\Xi^-}\)                                    & $\mbox{\small -0.677}$ &\(\mbox{\small -0.711}\) & \(\mbox{\small -0.745}\) & \(\mbox{\small -0.779}\) & \(\mbox{\small -0.6507}\)\\ \hline\hline
\(R\)                                              & $\mbox{\small 1.345}$  &\(\mbox{\small 1.623}\)  & \(\mbox{\small 1.954}\)  & \(\mbox{\small 2.356}\)  & \(\mbox{\small 1.676}\)  \\ \hline
 \end{tabular}
\end{center}
\caption{Quark model results for the magnetic moments, including
only octet loops. \label{quark_octet}}
\end{table}
\begin{table}
\begin{center}
  \begin{tabular}{|c|c|c|c|c|c|c|c|c||c||c}\hline
\small\( \Lambda  [\mbox{\footnotesize MeV}]\)     &\(\mbox{\small 300}\)    & \(\mbox{\small 400}\)    & \(\mbox{\small 500}\)    & \(\mbox{\small 600}\)    & \(\mbox{\small 700}\)    \\ \hline\hline
      \( \mu_u\)                                   &\(\mbox{\small 1.687}\)  & \(\mbox{\small 1.561}\)  & \(\mbox{\small 1.412}\)  & \(\mbox{\small 1.247}\)  & \(\mbox{\small 1.068}\)  \\ \hline
      \( \mu_d\)                                   &\(\mbox{\small -0.825}\) & \(\mbox{\small -0.718}\) & \(\mbox{\small -0.598}\) & \(\mbox{\small -0.468}\) & \(\mbox{\small -0.330}\) \\ \hline
      \( \mu_s\)                                   &\(\mbox{\small -0.590}\) & \(\mbox{\small -0.562}\) & \(\mbox{\small -0.522}\) & \(\mbox{\small -0.469}\) & \(\mbox{\small -0.407}\) \\ \hline\hline
      \(\mu_p\)                                    &\(\mbox{\small 2.793}\)  & \(\mbox{\small 2.793}\)  & \(\mbox{\small 2.793}\)  & \(\mbox{\small 2.793}\)  & \(\mbox{\small 2.793}\)  \\ \hline
      \(\mu_n\)                                    &\(\mbox{\small -1.913}\) & \(\mbox{\small -1.913}\) & \(\mbox{\small -1.913}\) & \(\mbox{\small -1.913}\) & \(\mbox{\small -1.913}\) \\ \hline
      \(\mu_{\Lambda}\)                            &\(\mbox{\small -0.613}\) & \(\mbox{\small -0.613}\) & \(\mbox{\small -0.613}\) & \(\mbox{\small -0.613}\) & \(\mbox{\small -0.613}\) \\ \hline
\(\mu_{\Sigma^+}\)                                 &\(\mbox{\small 2.607}\)  & \(\mbox{\small 2.563}\)  & \(\mbox{\small 2.519}\)  & \(\mbox{\small 2.475}\)  & \(\mbox{\small 2.434}\)  \\ \hline
\(\mu_{\Lambda\Sigma^0}\)                          &\(\mbox{\small 1.610}\)  & \(\mbox{\small 1.601}\)  & \(\mbox{\small 1.592}\)  & \(\mbox{\small 1.587}\)  & \(\mbox{\small 1.583}\)  \\ \hline
\(\mu_{\Sigma^-}\)                                 &\(\mbox{\small -1.018}\) & \(\mbox{\small -0.964}\) & \(\mbox{\small -0.903}\) & \(\mbox{\small -0.837}\) & \(\mbox{\small -0.767}\) \\ \hline
\(\mu_{\Xi^0}\)                                    &\(\mbox{\small -1.398}\) & \(\mbox{\small -1.379}\) & \(\mbox{\small -1.361}\) & \(\mbox{\small -1.345}\) & \(\mbox{\small -1.332}\) \\ \hline
\(\mu_{\Xi^-}\)                                    &\(\mbox{\small -0.526}\) & \(\mbox{\small -0.540}\) & \(\mbox{\small -0.550}\) & \(\mbox{\small -0.555}\) & \(\mbox{\small -0.554}\) \\ \hline\hline
\(R\)                                              &\(\mbox{\small 0.640}\)  & \(\mbox{\small 0.821}\)  & \(\mbox{\small 1.014}\)  & \(\mbox{\small 1.205}\)  & \(\mbox{\small 1.383}\)  \\ \hline
 \end{tabular}
\end{center}

\begin{center}
   \begin{tabular}{|c|c|c|c|c|c|c|c|c||c||c}\hline
\small\( \Lambda  [\mbox{\footnotesize MeV}]\)     & $\mbox{\small 800}$    &\(\mbox{\small 900}\)    & \(\mbox{\small 1000}\)   & \(\mbox{\small 1100}\)   &   \small {exp.}          \\ \hline\hline
      \( \mu_u\)                                   & $\mbox{\small 0.878}$  &\(\mbox{\small 0.680}\)  & \(\mbox{\small 0.474}\)  & \(\mbox{\small 0.263}\)  & \(\mbox{\small }\)       \\ \hline
      \( \mu_d\)                                   & $\mbox{\small -0.188}$ &\(\mbox{\small -0.041}\) & \(\mbox{\small 0.110}\)  & \(\mbox{\small 0.263}\)  & \(\mbox{\small }\)       \\ \hline
      \( \mu_s\)                                   & $\mbox{\small -0.335}$ &\(\mbox{\small -0.255}\) & \(\mbox{\small -0.168}\) & \(\mbox{\small -0.074}\) & \(\mbox{\small }\)       \\ \hline\hline
      \(\mu_p\)                                    & $\mbox{\small 2.793}$  &\(\mbox{\small 2.793}\)  & \(\mbox{\small 2.793}\)  & \(\mbox{\small 2.793}\)  & \(\mbox{\small 2.793}\)  \\ \hline
      \(\mu_n\)                                    & $\mbox{\small -1.913}$ &\(\mbox{\small -1.913}\) & \(\mbox{\small -1.913}\) & \(\mbox{\small -1.913}\) & \(\mbox{\small -1.913}\) \\ \hline
      \(\mu_{\Lambda}\)                            & $\mbox{\small -0.613}$ &\(\mbox{\small -0.613}\) & \(\mbox{\small -0.613}\) & \(\mbox{\small -0.613}\) & \(\mbox{\small -0.613}\) \\ \hline
\(\mu_{\Sigma^+}\)                                 & $\mbox{\small 2.395}$  &\(\mbox{\small 2.359}\)  & \(\mbox{\small 2.326}\)  & \(\mbox{\small 2.296}\)  & \(\mbox{\small 2.458}\)  \\ \hline
\(\mu_{\Lambda\Sigma^0}\)                          & $\mbox{\small 1.582}$  &\(\mbox{\small 1.583}\)  & \(\mbox{\small 1.586}\)  & \(\mbox{\small 1.591}\)  & \(\mbox{\small 1.61}\)   \\ \hline
\(\mu_{\Sigma^-}\)                                 & $\mbox{\small -0.695}$ &\(\mbox{\small -0.621}\) & \(\mbox{\small -0.545}\) & \(\mbox{\small -0.468}\) & \(\mbox{\small -1.16}\)  \\ \hline
\(\mu_{\Xi^0}\)                                    & $\mbox{\small -1.322}$ &\(\mbox{\small -1.314}\) & \(\mbox{\small -1.308}\) & \(\mbox{\small -1.304}\) & \(\mbox{\small -1.25}\)  \\ \hline
\(\mu_{\Xi^-}\)                                    & $\mbox{\small -0.547}$ &\(\mbox{\small -0.536}\) & \(\mbox{\small -0.520}\) & \(\mbox{\small -0.500}\) & \(\mbox{\small -0.6507}\)\\ \hline\hline
\(R\)                                              & $\mbox{\small 1.540}$  &\(\mbox{\small 1.672}\)  & \(\mbox{\small 1.776}\)  & \(\mbox{\small 1.854}\)  & \(\mbox{\small 1.676}\) \\ \hline
 \end{tabular}
\end{center}
\caption{Quark model results for the magnetic moments
including octet and decuplet loops.
\label{quark_decuplet}}
\end{table}

Again we see that a good phenomenological description is produced, with only these ingredients.
The overall results provide a good description of the baryon moments. For many values
of $\Lambda$ the quark moments are of a reasonable size. Note that since there can be chiral
renormalizations of the individual quark moments, the quark moments need not satisfy
the usual relations such as $\mu_d/\mu_u =-1/2$. However, for most values of $\Lambda$ the
modifications appear not to be large. Note, however, that for larger values of $\Lambda$
including decuplet loops, the modifications of the usual relations are over 100\%,
casting doubt on the phenomenological utility of these cases.

If we return to the $\Sigma^+$ puzzle we see that the chiral loops
have indeed introduced a new component to the magnetic moments
that is effective in changing the ratio. The results are given in the bottom lines
of Table~\ref{quark_octet} and
Table~\ref{quark_decuplet}. For example, from the tables
 with $\Lambda = 700$~MeV, we find
\begin{eqnarray}
R &=& 1.345 ~~~~~~~~~{\rm octet~ only  } \nonumber \\
  &=& 1.383 ~~~~~~~~~~~{\rm with~ decuplet}
\end{eqnarray}
The agreement with experiment is acceptable given that we are not
including any further sources of SU(3) breaking. The basic
message is clear. For reasonable values of the cutoff, the chiral
loops readily address some of the puzzles of the quark model. The
loops give a mechanism that shifts the $\Sigma^+$ ratio in the
right direction without destroying the general level of agreement
between theory and experiment. A conceptually simple way to
understand the baryon magnetic moments then starts with quark
moments and augments these with modest corrections from the chiral loops. The
$\Sigma^+$ puzzle is likely an indicator of the presence of
the chiral loops.

\section{Conclusions}
We have found that the long distance part of the kaon loop
calculation is a relatively small fraction of the result that has
been reported as the consequence of chiral perturbation theory~\cite{hemm}.
Because the chiral theory is meant to be an effective field theory
valid only in the long distance regime, the short distance parts
of the calculation are spurious and are not true consequences of
QCD. Chiral loops, when treated including all scales, produce
corrections that are so large that they disrupt standard
phenomenology. However, the need for such loops is clear and
they should not be totally discarded.
We therefore propose to keep only the long distance
contributions as model-independent the consequences of the kaon loop.

This work resolves the discrepancy between the chiral studies and
dispersive calculations of the same quantity, which also suggest
a considerably smaller strange radius. If we look carefully at the
dispersive calculation, we conclude that the reason is the same as
found above---the kaon mass is large and does not contribute to
the low energy (and therefore trustworthy) component of the
dispersive integral.

\section*{Acknowledgements}
This manuscript has been produced for a memorial volume for Dubravko Tadi$\acute{\rm c}$.
During his career Dubravko worked
on various aspects of the chiral quark model, and had a deep interest in
quarks and chiral physics.
We would like to thank Bastian Kubis, Martin Moj\v{z}i\v{s} and Thomas
Mannel for many useful discussions.
This work was supported in part by the National Science Foundation
under award PHY02-44801 and by
ERDF funds from the European Commission. This work was also supported
by the ``Studienstiftung des deutschen Volkes'' (Germany) and the
``Schweizerischer Nationalfond'' (Switzerland).

\end{document}